\documentclass[fdp,fleqn]{w-art}
\usepackage{times}
\usepackage{w-thm}
\def\AEF{A.E. Faraggi}

\def\NPB#1#2#3{\emph{ Nucl.\ Phys.}\/ B {\bf #1}, #3 (#2)}
\def\PLB#1#2#3{\emph{ Phys.\ Lett.}\/ B {\bf #1}, #3 (#2)}
\def\PRD#1#2#3{\emph{ Phys.\ Rev.}\/ D {\bf #1}, #3 (#2)}
\def\PRL#1#2#3{\emph{ Phys.\ Rev.\ Lett.}\/ {\bf #1}, #3 (#2)}

\def\EPJC#1#2#3{\emph{Eur.\ Phys.\ Jour.}\/ C {\bf #1}, #3 (#2)}

\usepackage[]{graphicx}
\begin{document}
\DOIsuffix{theDOIsuffix}
\Volume{55}
\Issue{1}
\Month{01}
\Year{2007}
\pagespan{1}{}



\title[MSHSM]{MSHSM -- Minimal Standard Heterotic String Models}


\author[A.E. faraggi]{Alon E. Faraggi\inst{1}%
  \footnote{Corresponding author\quad E-mail:~\textsf{alon.faraggi@liv.ac.uk}, 
            Phone: +00\,44\,0151\,7943774, 
            Fax: +00\,44\,0151\,7943784}}
\address[\inst{1}]{Department of
Mathematical Sciences, University of Liverpool, Liverpool L69 7ZL, UK}
\begin{abstract}
An overview of old and new results in studies of the quasi--realistic free fermionic
models is presented, which include the recent discovery of exophobic string vacua and
reproduction of the Higgs--matter splitting mechanism in a corresponding 
orbifold construction. 

\end{abstract}
\maketitle                   






Contemporary terrestrial experiments,
and extra--terrestrial observatories, indicate that the 
Standard $SU(3)_C\times SU(2)_L\times U(1)_Y$ Model of the 
gauge interactions, augmented with three families of quarks and 
leptons, correctly accounts for the extracted data. The remaining
piece, the scalar Higgs field and the mechanism of electroweak 
symmetry breaking, awaits the data and final judgement of the 
LHC experiment. Post 1998 data from astrophysical and reactor
neutrino experiments also indicate that neutrinos are 
massive, which suggests augmentation of each Standard Model generation
with a Standard Model singlet. 

Perhaps its most tantalising property, is the fact that the Standard Model 
matter and gauge sectors can be embedded into Grand Unified Theories. 
Most appealing in this context is the embedding into $SO(10)$, that 
accomodates each matter generation into a single spinorial 
16 representation. The gauge charges of the matter states are experimental
observables and the embedding into $SO(10)$ reduces these number 
of observables from $54=3\times 3\times 6$, to just one, namely the 
number of generations. This could of course be a cruel accident, 
but it is the best clue we have for the fundamental origins of the
Standard Model. 

Additional evidence for the validity of grand unification exists in the 
logarithmic running of the Standard Model parameters, which is compatible 
with the data in the gauge and heavy matter sectors. The 
Higgs sector receives quadratic corrections from the cutoff scale, and
restoration of its logarithmic evolution requires the introduction of a 
new symmetry -- {\it e.g.} supersymmetry. Additionally, proton longevity
and neutrino masses suggest that the unification scale is vastly separated 
from the electroweak scale. Indeed, proton stability is an accidental feature
of the Sandard Model at the renormalizable level. Non--renormalizable 
operators, suppressed by an effective cutoff scale, which must exist because 
the Standard Model is incomplete, will, in general, induce fast proton
decay. The proton lifetime is therefore another fact, set in our bones,
that hints at a large unification scale. 

Gran unification, however, cannot be the end of the story. The flavour
data is not explained in this context. The choice of gauge and matter
sectors is also ad hoc. Most importantly, gravity, is not included in
this framework. Furhermore, non--conformal quantum field theories and
general relativity are incompatible at a fundamental level. String theory
is a unique extension of the Standard Model in the sense that
while providing a consistent theory of quantum gravity, it also
gives rise to the gauge and matter structures that appear in the 
Standard Model. It provides a unifying framework for all
the basic particle constituents that appear in nature, 
and enables the development of a phenomenological approach 
to quantum gravity. Furthermore, string theory is predictive. 
It predicts that the rank of the gauge interactions is extended,
and should be, for example, below 22 in the heterotic string.

Given that there are five different theories in ten dimensions, 
which are all believed to be connected by various duality
transformations, a natural question is which of these theories
should be used as a starting point for phenomenology. The Standard
Model data motivates the hypothesis that the perturbative 
string vacuum should accommodate the canonical $SO(10)$ embedding
of its spectrum. Additionally, a perturbative string vacuum 
should contain three chiral generations and a pair 
of Higgs doublets with potentially viable Yukawa couplings. 
Among the five perturbative string theories the only one that 
which is compatible with these requirements is the $E_8\times
E_8$ heterotic--string, as it is the only one that produces 
spinorial representations in the perturbative massless spectrum. 
It may well be that exploring other properties of the 
fundamental vacuum will necessitate utilising a different 
perturbative limit of the underlying non--perturbative theory. 
The ultimate question in this regard is how are the different 
limits connected, and how can we be guaranteed that we are exploring
different limits of a single nonperturbative vacuum. 

The quasi--realistic free fermionic heterotic--string models
were constructed in the late eighties and early nineties
\cite{fsu5,fny,alr,nahe,eu,top,cslm,cfn,cfs,su421,fmt,cfmt}.
They provide a
concrete framework to study many of the issues that pertain to the
phenomenology of the Standard Model
and grand unification. A few highlights of studies in the free
fermionic standard--like models are listed below:

\begin{itemize}
\item[{$\bullet$}] { Top quark mass $\sim$ 175--180GeV  \cite{top}}
\item[{$\bullet$}] { Generation mass hierarchy \cite{fmm}}
\item[{$\bullet$}] { CKM mixing  \cite{ckm,fh}}
\item[{$\bullet$}] { Stringy seesaw mechanism \cite{seesaw,fh2,seesawII}}
\item[{$\bullet$}] { Gauge coupling unification \cite{gcu,df}}
\item[{$\bullet$}] { Proton stability \cite{ps}}
\item[{$\bullet$}] { Squark degeneracy \cite{fp2,fv}}
\item[{$\bullet$}] { Minimal Standard Heterotic String Model (MSHSM) 
                                                                \cite{cfn}}
\item[{$\bullet$}] { Moduli fixing \cite{modulifixing}}
\item[{$\bullet$}] { 
Classification \cite{nooij, classification} \&
spinor--vector duality \cite{spduality,tristan}}
\end{itemize}

The fermionic formulation was developed in the mid--eighties \cite{abk,klt}.
The equivalence of bosons and fermions is two dimensions 
entails that a model constructed
using the fermionic approach corresponds to a model
constructed using the bosonic approach in which the
target--space is compactified on a six dimensional internal manifold.
The free fermionic formalism corresponds to using
a free bosonic formalism in which the radii of the internal dimensions 
are fixed at a special point in the compact space.
Deformation from the special point in the moduli space are parametrized 
in terms of world--sheet Thirring interactions among the world--sheet 
fermions \cite{egrs}.
The simplicity of the free fermionic formalism entails that
the string consistency constraints are solved in terms of
the world--sheet free fermion transformation properties on the string
world--sheet, which are encoded in sets of basis vectors and
one--loop GSO projection coefficients among the basis vectors.
The formalism to extract the physical spectrum and superpotential
interaction terms are also straightforward. The simplest free fermionic
constructions correspond to a $Z_2\times Z_2$ orbifold of a six dimensional
toroidal manifold, augmented with discrete Wilson lines that are needed
to break the $SO(10)$ GUT symmetry.

Perhaps, the most tantalising achievement is the successful calculation of the
top quark mass \cite{top},
which was obtained several years prior to the experimental
discovery. This calculation demonstrats how string theory enables the
calculation of the fermion--scalar Yukawa couplings in terms of the
unified gauge coupling. Furthermore, the string  models offer
an explanation for the hierarchical mass splitting between the
top and bottom quarks.
The top quark Yukawa coupling is obtained at the cubic level
of the superpotential and is of order one, whereas the Yukawa 
couplings of the lighter quarks and leptons are obtained from
nonrenormalizable operators that are suppressed relative to the 
leading cubic level term. Thus, only the top quark mass is
characterised by the electroweak scale and the masses of the lighter
quarks and leptons are naturally suppressed compared to it. 
As the heavy generation Yukawa couplings
are obtained at low orders in the superpotential, the calculation of these
Yukawa couplings is robust and is common to a large class of models. 
Further analysis of fermion masses shows that
quasi--realistic fermion mass textures arise for reasonable
choices of supersymmetric flat directions \cite{ckm}. Issues like left--handed
neutrino masses \cite{fh2}, gauge coupling unification \cite{gcu,df}, 
proton stability \cite{ps} and squark 
degeneracy \cite{fp2,fv}
were studied in concrete quasi--realistic free fermionic string
models and for detailed solutions of the supersymmetric flat direction 
constraints. 
It was also demonstrated in ref. \cite{cfn} that the free fermionic
heterotic string vacua give rise to models that produce in the 
observable charged sector below the string 
unification scale solely the matter
spectrum of the minimal supersymmetric standard model.
Such models are dubbed Minimal Standard Heterotic 
String Models (MSHSM). The free fermionic models also provide important
clues to the problem of moduli fixing in string theory \cite{modulifixing}.
They 
highlight the fact that string theory may utilize geometrical structures
that do not have a classical correspondence. Primarily, they allow boundary
conditions that distinguish between the left-- and right--moving
coordinates of the six dimensional compactified space. 
Such boundary conditions necessarily lead to the projection
of the moduli fields associated with the extra internal coordinates.
The free fermionic models have also been instrumental in recent years to
unravel a new duality symmetry under the exchange of spinor and vector 
representations of the GUT group \cite{spduality, tristan}. 
The free fermionic formalism is also adaptable to a computerised 
classification, of $Z_2\times Z_2$ orbifolds with symmetric
shifts \cite{nooij, classification}.
A generic property of string vacua is the existence of fractionally charged 
states, due to the breaking of the $SO(10)$ GUT symmetry by discrete Wilson 
lines. The lowest state among those is stable due to electric charge 
conservation. Experimental constriants severely constrain the mass
and abundance of such such particles. Using the free fermionic
classification method we recently constructed examples of Pati--Salam
string models that are completely free of massless exotics \cite{exophobic}. 

The quasi--realistic free fermion models were constructed in the
late eighties and early nineties, and continue to be represent
viable string vacua that merit deeper exploration into their properties.
In particular, their relation to $Z_2\times Z_2$ orbifolds is of
vital importance. One should note that the different formalisms of
string compactifications are related and provide complementary insight
to the properties of the underlying vacua. So, for example,
while the free fermionic approach provides easy access to the
massless spectra and potential, and is adaptable to a computerised 
classification, orbifold representation may be more amenable to study
moduli dynamics and vacuum selection. In this vein, it is vital to
establish a detailed dictionary between the two laguages.
While each language may be well understood on its own,
writing a detailed dictionary between the two is often non--trivial. 

In this talk I highlight how this correspondence can be established
for two key properties of the models, and is indeed found to be highly
non--trivial. The first is with respect to the $Z_2\times Z_2$
orbifold action on the $SO(12)$ lattice at the free fermionic point.
It is well known that a $Z_2\times Z_2$ orbifold of a generic six
dimensional torus produces 48 fixed points, whereas $Z_2\times Z_2$
of the $SO(12)$ lattice produces 24 fixed points. The 48 points can
be reduced to 24 by adding a freely acting shift, under which
pairs of fixed points are identified. Now one can add different
freely acting shifts to this effect,
and the question is which one reproduces the free fermionic model.
Namely, it is not sufficient
to examine the massless spectrum for this purpose, but one must also 
establish the 
correspondence of the massive spectrum. String theory affords even more
interesting possibilities
because, in addition to the ordinary Kaluza--Klein momentum modes, it
also gives rise to 
space--like winding modes. Therefore, one can add freely acting shifts
that shift the momenta,
winding, as well as asymmetric shifts that act on both the momenta and
winding and are therefore
intrinsicallly stringy! It turms out that the shifts that reproduce
the partition function at the 
free fermionic point are uniquely of the third kind.

The second property, related to the free
fermion--orbifold correspondence, illustrates the utility of the free fermion 
method in accessing classes of string vacua that are 
obscure from the point of view of the orbifold contruction.
In the orbifold models one typically starts with the
$E_8\times E_8$ heterotic--string compactified to
four dimensions and subsequently break one $E_8$ gauge factor
by the orbifold twistings and discrete Wilson lines.
In contrast the quasi--realistic free fermionic models
start with  $SO(16)\times SO(16)$, where the reduction from
$E_8\times E_8$ to $SO(16)\times SO(16)$ is realised in
terms of a Generalised GSO (GGSO) phase in the partition function.
One can then show that the partition function at the free fermionic 
point is generated by starting with the partition function of the
$E_8\times E_8$ and acting on it with a freely acting $Z_2\times
Z_2$ twist given by 
$Z_2^a\times~Z_2^b~:~= 
(-1)^{F_{\xi^1}}\delta_1 ~\times~ (-1)^{F_{\xi^2}}\delta_1 \,$
with 
$\delta_1 X^9 = X_9 +\pi R_9~,$
where $F_{\xi^{1,2}}$ are fermion numbers acting on the first and
second $E_8$ factors,
and $\delta_9$ is a mod 2 shift in one internal direction 
\cite{parti}.
The action of the $Z_2\times Z_2$
freely acting twist reduces the $E_8\times E_8$ symmetry to
$SO(16)\times SO(16)$. 
The next step in the construction employs a $Z_2^c$ twist
that acts on the internal 
coordinates and produces 16 fixed points. The gauge symmetry is broken to 
$E_7\times SU(2)$, and $SO(12)\times SO(4)$,  in each case, respectively.
The matter and Higgs states in the first case are in the 56
representation of $E_7$,
which breaks as $56 = (32,1) + (12,2)$ under $SO(12)\times SU(2)$, where
the $32$ spinorial, and the $12$ vectorial, representations of $SO(12)$
contain the matter and Higgs representations of the Standard Model,
respectively. 
Hence, the freely acting $Z_2^a\times Z_2^b$ twists induce a string
Higgs--matter
splitting mechanism, similar to the string doublet--triplet splitting
mechanism \cite{ps}. 
However, it turns out that simply adding the $Z_2^c$ twist, {\it i.e.} taking
$[Z_+/(Z_2^a\times Z_2^b)]/Z_2^c$, where $Z_+$ is the $E_8\times E_8$ 
partition function, projects the twisted massless spinorial representations
and keeps the vectorials. 
To restore the matter representations one needs to analyse
the full $Z_+/(Z_2^a\times Z_2^b\times Z_2^c)$ partition and take into account 
the seven independent modular orbits (seven discrete torsions) \cite{cft}. A
non--trivial proposition indeed! Now, one should note that this is merely 
one example. The free fermionic classification accesses $2^5$ vacua at this 
level, and by breaking the hidden $SO(16)$ factor to $SO(8)\times SO(8)$
this is increased to $2^9$. We therefore note that the free fermionic
approach provides a superior tool to easily access
large classes of string vacua.
The bosonic formalism provides a more intuitive approach to the string moduli
and hopefully to the dynamics that fix the string vacuum.

\begin{acknowledgement}
This work is supported in part by the 
STFC under grant ST/G00062X/1 and the EU under
contract MRTN-CT-2006-035863. 

\end{acknowledgement}


\begin{thebibliography}{10}

\bibitem{fsu5} I.\ Antoniadis, J.\ Ellis, J.\ Hagelin and D.V.\ Nanopoulos
                \PLB{231}{1989}{65}.

\bibitem{fny} A.E.\ Faraggi, D.V.\ Nanopoulos and K.\ Yuan,
                                                 \NPB{335}{1990}{347}.

\bibitem{alr}	I. Antoniadis. G.K. Leontaris and J. Rizos,
                                \PLB{245}{1990}{161}.

\bibitem{nahe} A.E. Faraggi and D.V. Nanopoulos, \PRD{48}{1993}{3288}.

\bibitem{eu} \AEF, \PLB{278}{1992}{131}; \NPB{403}{1993}{101}.

\bibitem{top} \AEF, \PLB{274}{1992}{47}; \PLB{339}{1994}{223}.

\bibitem{cslm}	A.E. Faraggi, \NPB{387}{1992}{239}.


\bibitem{cfn} G.B.\ Cleaver, \AEF~ and D.V.\ Nanopoulos,
                       \PLB{455}{1999}{135}. 

\bibitem{cfs} G.B.\ Cleaver, A.E.\ Faraggi and C.\ Savage,
                                \PRD{63}{2001}{066001}. 
 
\bibitem{su421} G. Cleaver, A.E. Faraggi and S.~Nooij, \NPB{672}{2003}{64}. 

\bibitem{fmt} \AEF, E. Manno and C. Timirgaziu, \EPJC{50}{2007}{701}.

\bibitem{cfmt} G. Cleaver, \AEF, E. Manno and C. Timirgaziu,
\PRD{78}{2008}{046009}.

\bibitem{abk}  I.~Antoniadis, C.~P. Bachas, and C.~Kounnas,
\NPB{289}{1987}{87}.
 
\bibitem{fmm} \AEF, \NPB{407}{1993}{57}.

\bibitem{ckm} I. Antoniadis, J. Rizos and K. Tamvakis, \PLB{278}{1992}{257}.

\bibitem{fh} \AEF~and E. Halyo, \PLB{307}{1993}{305}; \NPB{416}{1994}{63}.

\bibitem{seesaw} I. Antoniadis, J. Rizos and K. Tamvakis, \PLB{279}{1992}{281}.

\bibitem{fh2} \AEF~and E. Halyo, \PLB{307}{1993}{311}.

\bibitem{seesawII} C. Coriano and \AEF, \PLB{581}{2004}{99}.

\bibitem{gcu} \AEF, \PLB{302}{1992}{202}. 

\bibitem{df} K.R. Dienes and \AEF, \PRL{75}{1995}{2646};
                                           \NPB{457}{1995}{409}.             
\bibitem{ps} \AEF, \NPB{428}{1994}{111}; 
                   \PLB{499}{2001}{147};
                   \PLB{520}{2001}{337}.

\bibitem{fp2} \AEF~ and J.C. Pati, \NPB{526}{1998}{21}.

\bibitem{fv} \AEF~ and O. Vives, \NPB{641}{2002}{93}.

\bibitem{modulifixing} \AEF, \NPB{728}{2005}{83}. 

\bibitem{nooij} \AEF, C. Kounnas, S.E.M. Nooij and J. Rizos, 
                                   \NPB{695}{2004}{41}.

\bibitem{classification}   A.E.Faraggi, C.~Kounnas and J.~Rizos,
                               \PLB{648}{2007}{84}; 

\bibitem{spduality} A.E.Faraggi, C.~Kounnas and J.~Rizos,
                                \NPB{774}{2007}{208};
                                \NPB{799}{2008}{19}.

 \bibitem{tristan} T. Catelin-Jullien, \AEF, C. Kounnas and J. Rizos, 
                      \NPB{812}{2009}{103}.

\bibitem{klt} H. Kawai, D.C. Lewellen, and S.H.-H. Tye, \NPB{288}{1987}{1}.

\bibitem{egrs} S. Elitzur, E. Gross, E. Rabinovici and N. Seiberg,
  \NPB{283}{1987}{413}.

\bibitem{exophobic} B. Assel, K. Christodoulides, A.E. Faraggi, C. Kounnas and 
                       John Rizos, \PLB{683}{2010}{306}. 

\bibitem{parti} \AEF and M. Tsulaia, \PLB{683}{2010}{314}.

\bibitem{hms} \AEF, \EPJC{49}{2007}{803}.  

\bibitem{cft} C. Angelantonj, \AEF~ and M. Tsulaia, paper in preparation. 
\end{thebibliography}
\end{document}